\documentstyle[12pt]{article}

\catcode`\@=11
\long\def\@makefntext#1{ 
\protect\noindent \hbox to 3.2pt {\hskip-.9pt
$^{{\ninerm\@thefnmark}}$\hfil}#1\hfill} 

\def\thefootnote{\fnsymbol{footnote}}
 \def\@makefnmark{\hbox to 0pt{$^{\@thefnmark}$\hss}}  

\def\ps@myheadings{\let\@mkboth\@gobbletwo
\def\@oddhead{\hbox{} 
\rightmark\hfil\ninerm\thepage}
\def\@oddfoot{}\def\@evenhead{\ninerm\thepage\hfil 
\leftmark\hbox{}}\def\@evenfoot{}
\def\sectionmark##1{}\def\subsectionmark##1{}}

\begin{document}

\newcommand{\symbolfootnote}{\renewcommand{\thefootnote}
	{\fnsymbol{footnote}}}
\renewcommand{\thefootnote}{\fnsymbol{footnote}}
\newcommand{\alphfootnote}
	{\setcounter{footnote}{0}
	 \renewcommand{\thefootnote}{\sevenrm\alph{footnote}}}

\newcounter{sectionc}\newcounter{subsectionc}\newcounter{subsubsectionc}
\renewcommand{\section}[1] {\vspace{0.6cm}\addtocounter{sectionc}{1}
\setcounter{subsectionc}{0}\setcounter{subsubsectionc}{0}\noindent
	{\bf\thesectionc. #1}\par\vspace{0.4cm}}
\renewcommand{\subsection}[1] {\vspace{0.6cm}\addtocounter{subsectionc}{1}
	\setcounter{subsubsectionc}{0}\noindent
	{\it\thesectionc.\thesubsectionc. #1}\par\vspace{0.4cm}}
\renewcommand{\subsubsection}[1]
{\vspace{0.6cm}\addtocounter{subsubsectionc}{1}
	\noindent {\rm\thesectionc.\thesubsectionc.\thesubsubsectionc.
	#1}\par\vspace{0.4cm}}
\newcommand{\nonumsection}[1] {\vspace{0.6cm}\noindent{\bf #1}
	\par\vspace{0.4cm}}

\newcounter{appendixc}
\newcounter{subappendixc}[appendixc]
\newcounter{subsubappendixc}[subappendixc]
\renewcommand{\thesubappendixc}{\Alph{appendixc}.\arabic{subappendixc}}
\renewcommand{\thesubsubappendixc}
	{\Alph{appendixc}.\arabic{subappendixc}.\arabic{subsubappendixc}}

\renewcommand{\appendix}[1] {\vspace{0.6cm}
        \refstepcounter{appendixc}
        \setcounter{figure}{0}
        \setcounter{table}{0}
        \setcounter{equation}{0}
        \renewcommand{\thefigure}{\Alph{appendixc}.\arabic{figure}}
        \renewcommand{\thetable}{\Alph{appendixc}.\arabic{table}}
        \renewcommand{\theappendixc}{\Alph{appendixc}}
        \renewcommand{\theequation}{\Alph{appendixc}.\arabic{equation}}
        \noindent{\bf Appendix \theappendixc #1}\par\vspace{0.4cm}}
\newcommand{\subappendix}[1] {\vspace{0.6cm}
        \refstepcounter{subappendixc}
        \noindent{\bf Appendix \thesubappendixc. #1}\par\vspace{0.4cm}}
\newcommand{\subsubappendix}[1] {\vspace{0.6cm}
        \refstepcounter{subsubappendixc}
        \noindent{\it Appendix \thesubsubappendixc. #1}
	\par\vspace{0.4cm}}

\def\abstracts#1{{
	\centering{\begin{minipage}{30pc}\tenrm\baselineskip=12pt\noindent
	\centerline{\tenrm ABSTRACT}\vspace{0.3cm}
	\parindent=0pt #1
	\end{minipage} }\par}}

\newcommand{\bibit}{\it}
\newcommand{\bibbf}{\bf}
\renewenvironment{thebibliography}[1]
	{\begin{list}{\arabic{enumi}.}
	{\usecounter{enumi}\setlength{\parsep}{0pt}
\setlength{\leftmargin 1.25cm}{\rightmargin 0pt}
	 \setlength{\itemsep}{0pt} \settowidth
	{\labelwidth}{#1.}\sloppy}}{\end{list}}

\topsep=0in\parsep=0in\itemsep=0in
\parindent=1.5pc

\newcounter{itemlistc}
\newcounter{romanlistc}
\newcounter{alphlistc}
\newcounter{arabiclistc}
\newenvironment{itemlist}
    	{\setcounter{itemlistc}{0}
	 \begin{list}{$\bullet$}
	{\usecounter{itemlistc}
	 \setlength{\parsep}{0pt}
	 \setlength{\itemsep}{0pt}}}{\end{list}}

\newenvironment{romanlist}
	{\setcounter{romanlistc}{0}
	 \begin{list}{$($\roman{romanlistc}$)$}
	{\usecounter{romanlistc}
	 \setlength{\parsep}{0pt}
	 \setlength{\itemsep}{0pt}}}{\end{list}}

\newenvironment{alphlist}
	{\setcounter{alphlistc}{0}
	 \begin{list}{$($\alph{alphlistc}$)$}
	{\usecounter{alphlistc}
	 \setlength{\parsep}{0pt}
	 \setlength{\itemsep}{0pt}}}{\end{list}}

\newenvironment{arabiclist}
	{\setcounter{arabiclistc}{0}
	 \begin{list}{\arabic{arabiclistc}}
	{\usecounter{arabiclistc}
	 \setlength{\parsep}{0pt}
	 \setlength{\itemsep}{0pt}}}{\end{list}}

\newcommand{\fcaption}[1]{
        \refstepcounter{figure}
        \setbox\@tempboxa = \hbox{\tenrm Fig.~\thefigure. #1}
        \ifdim \wd\@tempboxa > 6in
           {\begin{center}
        \parbox{6in}{\tenrm\baselineskip=12pt Fig.~\thefigure. #1 }
            \end{center}}
        \else
             {\begin{center}
             {\tenrm Fig.~\thefigure. #1}
              \end{center}}
        \fi}

\newcommand{\tcaption}[1]{
        \refstepcounter{table}
        \setbox\@tempboxa = \hbox{\tenrm Table~\thetable. #1}
        \ifdim \wd\@tempboxa > 6in
           {\begin{center}
        \parbox{6in}{\tenrm\baselineskip=12pt Table~\thetable. #1 }
            \end{center}}
        \else
             {\begin{center}
             {\tenrm Table~\thetable. #1}
              \end{center}}
        \fi}

\def\@citex[#1]#2{\if@filesw\immediate\write\@auxout
	{\string\citation{#2}}\fi
\def\@citea{}\@cite{\@for\@citeb:=#2\do
	{\@citea\def\@citea{,}\@ifundefined
	{b@\@citeb}{{\bf ?}\@warning
	{Citation `\@citeb' on page \thepage \space undefined}}
	{\csname b@\@citeb\endcsname}}}{#1}}

\newif\if@cghi
\def\cite{\@cghitrue\@ifnextchar [{\@tempswatrue
	\@citex}{\@tempswafalse\@citex[]}}
\def\citelow{\@cghifalse\@ifnextchar [{\@tempswatrue
	\@citex}{\@tempswafalse\@citex[]}}
\def\@cite#1#2{{$\null^{#1}$\if@tempswa\typeout
	{IJCGA warning: optional citation argument
	ignored: `#2'} \fi}}
\newcommand{\citeup}{\cite}

\def\fnm#1{$^{\mbox{\scriptsize #1}}$}
\def\fnt#1#2{\footnotetext{\kern-.3em
	{$^{\mbox{\sevenrm #1}}$}{#2}}}

\font\twelvebf=cmbx10 scaled\magstep 1
\font\twelverm=cmr10 scaled\magstep 1
\font\twelveit=cmti10 scaled\magstep 1
\font\elevenbfit=cmbxti10 scaled\magstephalf
\font\elevenbf=cmbx10 scaled\magstephalf
\font\elevenrm=cmr10 scaled\magstephalf
\font\elevenit=cmti10 scaled\magstephalf
\font\bfit=cmbxti10
\font\tenbf=cmbx10
\font\tenrm=cmr10
\font\tenit=cmti10
\font\ninebf=cmbx9
\font\ninerm=cmr9
\font\nineit=cmti9
\font\eightbf=cmbx8
\font\eightrm=cmr8
\font\eightit=cmti8

\newcommand{\be}{\begin{equation}}
\newcommand{\ee}{\end{equation}}
\newcommand{\bea}{\begin{eqnarray}}
\newcommand{\eea}{\end{eqnarray}}
\newcommand{\nen}{\nonumber \\ \relax}
\renewcommand{\theequation}{\arabic{equation}}
\newcommand{\seq}{\ =\ }
\newcommand{\pls}{\ +\ }
\newcommand{\mi}{\ -\ }
\newcommand{\cc}{{\cal C}}
\newcommand{\cm}{{\cal M}}
\newcommand{\co}{{\cal O}}
\newcommand{\dual}{^\star\!}
\newcommand{\prd}[1]{{\it Phys. Rev.} {\bf D#1}}
\newcommand{\ap}[1]{{\it Ann. Phys.} {\bf #1}}
\newcommand{\np}[1]{{\it Nucl. Phys.} {\bf B#1}}
\newcommand{\cmp}[1]{{\it Commun. Math. Phys.} {\bf #1}}
\newcommand{\ijmp}[1]{{\it Intl. J. Mod. Phys.} {\bf A#1}}
\newcommand{\jmp}[1]{{\it J. Math. Phys.} {\bf #1}}
\newcommand{\cqg}[1]{{\it Class. Quan. Grav.} {\bf #1}}
\newcommand{\mpl}[1]{{\it Mod. Phys. Lett.} {\bf A#1}}
\newcommand{\prl}[1]{{\it Phys. Rev. Lett.} {\bf #1}}
\newcommand{\pr}[1]{{\it Phys. Rev.} {\bf #1}}
\newcommand{\pl}[1]{{\it Phys. Lett.} {\bf #1B}}
\newcommand{\rmp}[1]{{\it Rev. Mod. Phys.} {\bf #1}}

\centerline{\tenbf MODULI SPACE COHOMOLOGY}
\baselineskip=22pt
\centerline{\tenbf AND WAVEFUNCTIONALS IN 3D QUANTUM GRAVITY\footnote{This is a
summary of a talk presented at the Seventh Marcel Grossmann Meeting on General
Relativity, Stanford University, Stanford, CA, USA, July 24 - 30, 1994}}
\vspace{0.8cm}
\centerline{\tenrm ROGER BROOKS}
\baselineskip=13pt
\centerline{\tenit Center for Theoretical Physics,}
\baselineskip=12pt
\centerline{\tenit Laboratory
for Nuclear Science}
\centerline{\tenit and Department of Physics,}
\centerline{\tenit Massachusetts Institute of Technology}
\baselineskip=12pt
\centerline{\tenit Cambridge, Massachusetts 02139 U.S.A.}
\vspace{0.9cm}
\abstracts{Wavefunctionals of three dimensional quantum gravity are extracted
from the 3D field theoretic analogs of the four dimensional Donaldson
polynomials.
Our procedure is generalizable to four and other dimensions.}
\vspace{0.3cm}
\leftline{CTP \# 2363, hep-th/9409146\hfill September 1994}
\vspace{0.5cm}
\twelverm   
\baselineskip=14pt
Since topological invariants are a subset of diffeomorphism invariants,
elucidation of their  role in  gravity is expected to be of (at least)
pedagogical value in constructing a quantum gravity theory.  We know how to
write down path integral expressions\cite{Wit(TYM)} for the Donaldson
polynomial invariants\cite{DonKro} of four dimensional manifolds.  Thus we will
investigate the role the topological gravity analogs of these invariants play
in ordinary quantum gravity.  The simplest setting to examine this is in three
dimensions where quantum gravity can be written as a $BF$ of Chern-Simons
topological field theory\cite{Wit(CSG)}.  In the interest of applications to
four and higher dimensions,  our discussion below will emphasize
generality\cite{BroLif}.

Consider a general field theory (GFT) with action $S_{GFT}[X]$ and a
topological quantum field theory (TQFT)\cite{TQFTrev} with action
\be
S_{TQFT}[X,Y]\seq S_{GFT}[X] ~+~ S_S[X,Y]\ \ ,\label{TQFTact}\ee
both on a manifold, $M$, with boundary, $\partial M=\Sigma$.
The set of fields $Y$ are akin to the super-partners of the GFT fields $X$.
Let $\co(X,Y)$ denote a product of observables in the TQFT whose total ghost
number is equal to the dimension of the moduli space the TQFT is defined for.
Construct the Hartle-Hawking wavefunctional (the notation $|_\Sigma$ means
restriction to $\Sigma$),
\be
\Psi[X|_{\Sigma},Y|_{\Sigma}]\seq\int [dX][dY] e^{-S_{TQFT}} \co(X,Y)\ \
,\label{PSI}\ee
in the TQFT by performing the indicated functional integrals with fixed
boundary values, $X|_\Sigma$ and $Y|_\Sigma$, for the fields.  Then set
$Y|_\Sigma=0$ in this TQFT wavefunctional to obtain a wavefunctional in the
GFT; {\it i.e.},
\be \Psi[X|_\Sigma,Y|_\Sigma]\Longrightarrow \Psi[X|_\Sigma]\ \ .\ee
Observe that the $\Psi[X|_\Sigma]$ differs considerably from the Hartle-Hawking
wavefunctional of the GFT.  In constructing it from Eq. (\ref{PSI}), the
functional integration over the ``superpartner" $Y$ yields a non-local
functional, $F[X]$, of the $X$ fields.  This means that generically,  the GFT
wavefunctional is of the form
\be
\Psi[X|_\Sigma]\seq \int [dX] F[X] e^{-S_{GFT}[X]}\ \ .\ee

The fact that the $\Psi[X|_\Sigma]$ so constructed satisfy the  constraints of
the GFT is seen as follows:
Given the  constraints, $\cc_{GFT}(X|_\Sigma)$, in the GFT, the set of
constraints in the TQFT includes the constraints
\be
\cc_{TQFT}(X|_\Sigma,Y|_\Sigma)\seq
\cc_{GFT}(X|_\Sigma)~+~\cc_S(X|_\Sigma,Y|_\Sigma)\ \ ,\ee
as may be seen from Eq. (\ref{TQFTact}).
Now, $\Psi[X|_{\Sigma},Y|_{\Sigma}]$ satisfies
\be
\cc_{TQFT}(X|_\Sigma,Y|_\Sigma)\Psi[X|_{\Sigma},Y|_{\Sigma}]\seq 0\ \ ,\ee
thus
$\cc_{GFT}(X|_\Sigma)\Psi[X|_{\Sigma},Y|_{\Sigma}]+
\cc_S(X|_\Sigma,Y|_\Sigma)\Psi[X|_{\Sigma},Y|_{\Sigma}]=0$.
After evaluating this at $Y|_{\Sigma}=0$ (which implies $\cc_S(X|_\Sigma,0)=0$
identically) and calling $\Psi[X|_\Sigma]\equiv \Psi[X|_\Sigma,0]$, we arrive
at
\be \cc_{GFT}(X|_\Sigma)\Psi[X|_\Sigma]\seq0\ \ .\ee
Namely, the wavefunctional so constructed satisfies the  constraints of the GFT
and is defined on the set of fields, $X|_\Sigma$, of the GFT only.

As an example of the procedure, let $M$ be a genus-$g$ handlebody, then we
formally find the quantum gravity wavefunctionals
\bea
\Psi[\omega;g]=&\int
[d\mu]_{SBF}~ \prod_{I=1}^{D_\cm}\delta(U_I(A)-
g_I(\omega))~ \prod_{J=1}^{D_\cm}\delta(\oint_{J}\psi)\times\nen
{}~&\hskip 45pt\times ~\prod_i\co_i(\phi,\psi,F)~e^{-S_{SBF}}\ \ ,\eea
after starting from the wavefunctions of super-$BF$ gauge theory\cite{TQFTrev}
(for flat $SO(3)$ connections).
Here $D_\cm=3(g-1)$ is the dimension of the moduli space of flat $SO(3)$
connections over the handlebody; $U_I(A)$ is the holonomy of the $SO(3)$
connection $A$ along each longitude; $\omega$ is a flat connection  -- with
holonomy $g_I(\omega)$ -- of the Riemann surface which bounds the handlebody;
$\psi$ is the ghost for the topological shift symmetry on $A$ and $\phi$ is a
scalar, secondary ghost. The delta functions enforce the boundary conditions on
the fields in  a gauge invariant manner.  The $\co_i$ are the observables
(elements of the equivariant cohomology) in the super-$BF$ theory with action
$S_{SBF}$ and functional measure $[d\mu]_{SBF}$.  This formal wavefunctional is
 to be computed in the semi-classical limit; this is known\cite{Wit(TYM)} to be
exact.  In this limit, for example, $\phi$ is replaced by an expression
involving $\psi$ and a Green's function\cite{BroLif}.

In the special case of three dimensional gravity, we can also obtain
wavefunctionals which solve the canonical constraints by ``almost" computing
correlation functions in a super-$BF$ gauge theory over
a {\it two-dimensional} genus-$g$ Riemann surface.  By ``almost" we mean
integrating
over all fields in the super-$BF$ gauge theory except for the connection; this
gives
\goodbreak
\bea
\Psi_{\vec n}[\omega;g] &\seq&  \hskip -18pt (-)^{n_3}\prod_i^{n_4}
Tr((\int_{y,\Sigma_{g}} G_\omega(x_i,y)[q(\omega(y)),\dual
q(\omega(y))])^2)~\times\nen
&\qquad\times&\prod_j^{n_3} \oint_{\gamma_j}Tr(\int_{y,\Sigma_{g}}
G_\omega(x_j,y)
[q(\omega(y)),\dual q(\omega(y))] q(\omega(x_j)))~\times\nen
&\qquad\times& \prod_k^{n_2}
\int_{\Sigma_{g}}Tr(  q(\omega)\wedge q(\omega))\ ,\label{CanWfcn}\eea
with $n_4+n_3+n_2=6(g-1)$;
where $G_\omega$ is the Green's function of the two-dimensional
scalar laplacian with connection $\omega$; $\int_{y,\Sigma_g}$ means the
integral over the genus-g Riemann surface, $\Sigma_g$, with coordinates $y$;
and the $q(\omega)$ form a $6(g-1)$-dimensional basis for $H^1(\Sigma_g,SO(3))$
and appear in a totally anti-symmetric combination.

In conclusion, starting with the Hilbert spaces of certain topological gravity
theories, we can construct subsets of wavefunctionals  for quantum gravity.
These are expected to form quantum gravity Hilbert sub-spaces   with an inner
product inherited from the topological gravity theory.  Work on the latter
issue is in progress.  Further details along with a sketch of our approach for
four dimensional quantum gravity is given in Ref.  [4].  Functional elaboration
of the relationship between $BF$ gauge theories and super-$BF$ gauge theories
may be found in Ref. [6].

\vspace{0.6cm}\noindent
	{\bf Acknowledgements}\par\vspace{0.4cm}
The author thanks Gilad Lifschytz for a collaboration\cite{BroLif} which was
the basis for this talk.
This work is supported in part by funds
provided by the
U. S. Department of Energy (D.O.E.) under cooperative agreement
\# DE-FC02-94ER40818.

\end{document}